\documentclass[journal,10pt]{IEEEtran}
\usepackage{amsfonts,amssymb}
\usepackage{amsmath}
\usepackage{array}
\usepackage{cite}
\usepackage{url}
\usepackage{xcolor}
\usepackage{cite,graphicx,amsmath,amssymb}
\usepackage{subfigure}
\usepackage{fancyhdr}
\usepackage{mdwmath}
\usepackage{mdwtab}
\usepackage{amsthm}
\usepackage{booktabs}
\usepackage{multirow}
\usepackage{longtable}
\usepackage{rotating}
\usepackage{makecell}
\usepackage[hidelinks]{hyperref}

\newtheorem{remark}{Remark}
\newtheorem{theorem}{Theorem}

\newtheorem{lemma}{Lemma}

\newtheorem{corollary}{Corollary}

\begin{document}

\title{Performance Analysis for Near-Field MIMO: Discrete and Continuous Aperture Antennas}

\author{Ziyi~Xie,~Yuanwei~Liu,~Jiaqi~Xu,~Xuanli~Wu,~and~Arumugam~Nallanathan

\thanks{Z. Xie and X. Wu are with the School of Electronics and Information Engineering, Harbin Institute of Technology, Harbin 150001, China (email: \{ziyi.xie, xlwu2002\}@hit.edu.cn). Y. Liu, J. Xu, and A. Nallanathan are with the School of Electronic Engineering and Computer Science, Queen Mary University of London, E1 4NS, U.K. (email: \{yuanwei.liu, jiaqi.xu,  a.nallanathan\}@qmul.ac.uk).}
}




\maketitle

\begin{abstract}
Performance analysis is carried out in a near-field multiple-input multiple-output (MIMO) system for both discrete and continuous aperture antennas. The effective degrees of freedom (EDoF) is first derived. It is shown that near-field MIMO systems have a higher EDoF than free-space far-field ones. Additionally, the near-field EDoF further depends on the communication distance. Based on the derived EDoF, closed-form expressions of channel capacity with a fixed distance are obtained. As a further advance, with randomly deployed receivers, ergodic capacity is derived. Simulation results reveal that near-field MIMO has an enhanced multiplexing gain even under line-of-sight transmissions. In addition, the performance of discrete MIMO converges to that of continuous aperture MIMO.
\end{abstract}


\section{Introduction}
Multiple-input multiple-output (MIMO) is the key technique in recent and future wireless networks for its capability of spatial multiplexing. Specifically, rich scattering in the propagation environment enables the high rank of channel matrix ${\bf H}$, and hence multiple parallel data streams can be simultaneously transmitted through the MIMO channel \cite{MIMOcapacity}. 

Utilizing higher frequency bands such as mmWave and terahertz is another evolution trend. Compared with signals in lower frequency bands, high-frequency signals suffer severe penetration loss resulting in shrunken coverage ranges. At the same time, due to the lack of diffraction, the communication in high frequency is mostly line-of-sight (LoS) \cite{mmWave}. The near-field propagation and LoS transmissions bring new features to MIMO. Since the effect of the spherical wave cannot be ignored, the path loss of MIMO subchannels is not uniform \cite{DaiNF,JiaqiNF}. In addition, mutual coupling among antennas should be considered \cite{Dardari}. In this case, the performance analysis of MIMO is different from that in far-field fading setups.

When recalling the channel capacity of discrete MIMO, if channel state information is known at both the transmitter and the receiver, the capacity can be achieved by the well-known water-filling power allocation
\begin{align}\label{eq: capacity exact}
	C = \max\limits_{P_i: \sum_{1 \le i \le \phi} P_i = P} \sum \limits_i \log_2 \left( 1+ \frac{P_i}{N_0} {s_i}^2 \right),
\end{align}
where $P$ is the transmit power, $N_0$ is the additive white Gaussian noise, $\phi$ is the rank of ${\bf H}$, and $s_1 \ge...\ge s_\phi > 0$ are the non-zero singular values of ${\bf H}$. In the fading environment, \cite{MIMOcapacity} obtained the closed-form performance by the full-rank feature of ${\bf H}$. For LoS MIMO in the near-field, however, it is difficult to further simplify \eqref{eq: capacity exact} as the exact closed form due to the sparsity of ${\bf H}$ \cite{LoSMIMO}. As \eqref{eq: capacity exact} is intractable, how MIMO parameters affect the capacity performance is unclear, and hence few guidelines on the system design are found. Motivated by the above, we propose a novel approach to derive the closed-form capacity approximations in the near-field, which have not been obtained in existing works.
To obtain the tractable approximation from \eqref{eq: capacity exact}, we introduce the concept of effective degrees of freedom (EDoF) from information theory \cite{defEDoF}. The channel capacity can be calculated as the sum capacity of multiple identical single-input single-output (SISO) channels, and EDoF is the number of SISO channels. Thus, we are able to evaluate the performance of MIMO by methods in simple SISO systems. 

In this letter, we investigate the performance of the near-field MIMO system for both {\it discrete and continuous aperture antennas}. {\it Firstly}, we derive closed-form expressions of EDoF which characterizes the number of equivalent SISO channels. {\it Secondly}, we express the capacity of MIMO exploiting the derived EDoF. On this basis, we derive the closed-form channel capacity conditioned on the predefined distance. The analytical expressions are consistent with the far-field results in the large distance limit. {\it Thirdly}, we consider randomly deployed receivers and derive the closed-form ergodic capacity by averaging over the random receiver locations. {\it Finally}, we evaluate the analytical results by simulations. Numerical results show that 1) without scattering clusters in the environment, near-field MIMO achieves up to a five-fold multiplexing gain in the considered setup; 2) the performance of the continuous-aperture MIMO is the upper bound of the discrete case.

{\it Notation:} Bold lowercase letters are used for vectors. Bold capital letters are used for sets and matrices. ${\bf A} \left(n,m \right)$ denotes the element at row $n$ and column $m$. $\left( \cdot \right)^*$ is Hermitian transposition. $\left\| \cdot \right\|_F$ denotes the Frobenius norm. ${\rm tr} \left( {\bf A} \right)$ stands for the trace of ${\bf A}$. ${\mathbb E} \left[ \cdot \right] $ is the expectation operator.

\section{Discrete Antennas}
\subsection{Channel Model}
Let us begin with conventional MIMO with discrete antennas. We consider a downlink MIMO system where both the transmitter and the receiver are equipped with a uniform linear array (ULA). 
The center of the transmitter is located at the origin of a three-dimensional plane $\mathbb{R}^3$, while the center of the receiver is at $(x_r,y_r,0)$ and $d = \sqrt{x_r^2 +y_r^2}$. The ULAs face each other and are parallel to the $z$-axis.

Let $N_t$ and $N_r$ denote the numbers of antennas of the transmitter and the receiver, respectively. The antenna spacing of all considered ULAs is $q$. Therefore, the position of the $m$-th antenna on the transmitter is ${\bf r}_{t_m} = \left(0,0, \left(m- \frac{N_t+1}{2}\right) q \right)$, and the $n$-th antenna on the receiver is at ${\bf r}_{r_n} = \left( x_r,y_r, \left(n- \frac{N_r+1}{2} \right) q \right)$. The channel matrix between the transmitter and the receiver is
\begin{align}
	\renewcommand\arraystretch{0.23}
	{\bf H} = \begin{bmatrix} G_{1,1} &\cdots & G_{1,N_t} \\ \vdots & \ddots & \vdots \\ G_{N_r,1} & \cdots & G_{N_r,N_t}  \end{bmatrix},
\end{align}
where $G_{n,m}$ is the channel gain between the $m$-th antenna of the transmitter and the $n$-th antenna of the receiver. In the free space\footnote{In this work, we focus on the radiating near-field, where we have $d \gg \lambda$ and $\lambda$ is the wavelength.}, $G_{n,m}$ can be expressed as follows 
\begin{align}\label{eq: G}
	G_{n,m} = \frac{l_e}{4\pi} \frac{\exp \left( -jk_0 D_{n,m} \right)}{D_{n,m}},
\end{align}
where $k_0 = \frac{2\pi}{\lambda}$ is the wavenumber. $D_{n,m} = \left|{\bf r}_{r_n} - {\bf r}_{t_m} \right| =  \sqrt{d^2 + \left( m-n - \frac{N_t - N_r}{2} \right)^2 q^2}$ is the distance between two points ${\bf r}_{r_n}$ and ${\bf r}_{r_m}$, and hence the channel gain is related to the Green's function, i.e., $G_{n,m} = l_e G \left({\bf r}_{r_n}, {\bf r}_{t_m} \right)$. Instead of modeling antenna elements at receivers as sizeless points, we consider that the aperture length of each antenna is $l_e \in (0,q]$.
The channel correlation matrix is given by
\begin{align}\label{eq: correl matrix}
	{\bf R} = {\bf H}^*{\bf H}.
\end{align}

For conventional discrete MIMO, $\phi$ is called degrees of freedom (DoF), which represents the maximum number of independent streams of information that can be transmitted at the high signal-to-noise ratio (SNR).  
When all MIMO subchannels $\{1,...,\phi\}$ are under similar channel conditions, i.e., $s_1 \approx ... \approx s_\phi$, the capacity in \eqref{eq: capacity exact} can be approximately simplified as $C = \phi \log_2 \left( 1 + \frac{P}{ \phi N_0} {s_1}^2 \right)$. 
In the near-field, however, the singular values of $\bf H$ vary from each other due to the effect of the spherical wave, and hence the capacity might be achieved without using all available subchannels. In the low power regime, we utilize the discussed concept of EDoF to characterize the MIMO capacity as follows
\begin{align}\label{eq: def ecapacity}
		C = \varepsilon \log_2 \left( 1 + \frac{\left\| {\bf H} \right\|_F P}{ \varepsilon^2 N_0} \right),
\end{align}
where $\left\| {\bf H} \right\|_F = {\rm tr} \left( {\bf R} \right)$ is the overall channel power and $\varepsilon \in \left[1, \phi\right]$ is the EDoF. The EDoF is approximated as \cite{defEDoF}
\begin{align}\label{eq: EDoF def}
	\varepsilon \approx \frac{ \left( {\rm tr} \left( {\bf R} \right) \right)^2}{ {\rm tr} \left(  {\bf R}^2 \right)}.
\end{align}

The detailed proof of \eqref{eq: EDoF def} is shown in Appendix A.

\subsection{EDoF Analysis}
For discrete MIMO, the EDoF expression \eqref{eq: EDoF def} can be further expressed as
\begin{align}
	\varepsilon_{\rm dis} = \frac{ \left| \sum_{m=1}^{N_t} {\bf R} \left( m, m \right) \right|^2 }{\sum_{m_1=1}^{N_t} \sum_{m_2=1}^{N_t} \left| {\bf R} \left( m_1, m_2 \right) \right|^2 }.
\end{align}

\begin{lemma}\label{lemma: EDOF d}
	The EDoF of the discrete MIMO system is
	\begin{align}
		\varepsilon _{\rm dis}  = \left(N_t N_r \right)^2 \bigg/ \sum_{m_1 = 1}^{N_t} \sum_{m_2 = 1}^{N_t}  \frac{\sin^2 \left( \frac{q^2k_0 \left(m_1 - m_2 \right) }{2d}  N_r \right)}{ \sin^2 \left( \frac{q^2k_0 \left( m_1 - m_2  \right)}{2d}   \right)}.
	\end{align}
\end{lemma}
\begin{IEEEproof}
	By employing the Taylor expansion $\sqrt{x +1} \cong 1+ \frac{x}{2}$ to calculate the phase term in \eqref{eq: G}, the element ${\bf R}(m_1,m_2)$ can be expressed as
	\begin{align}
		\left| {\bf R}(m_1,m_2) \right|^2  & \overset{(a)}{\approx} \left|\sum_{n=1}^{N_r} \frac{ {l_e}^2 e^  {-j \frac{k_0}{d} \left(z_{m_1} - z_{m_2} \right) z_n }}{(4\pi d)^2 } \right|^2 \nonumber \\
		&= \frac{{l_e}^4 }{(4\pi d)^4 } \left|\sum_{n=1}^{N_r} e^  {-j \frac{q^2k_0}{d} \left(m_1 - m_2 \right) n} \right|^2 \nonumber \\
		& \overset{(b)}{=} \frac{ {l_e}^4}{(4\pi d)^4 } \frac{\sin^2 \left( \frac{q^2k_0 \left(m_1 - m_2 \right) }{2d}  N_r \right)}{ \sin^2 \left( \frac{q^2k_0 \left(m_1 - m_2  \right)}{2d}   \right)},
	\end{align}
	where the approximation $(a)$ is valid for $d \ge 1.2 qN_t$ \cite{Emilconf}. This condition can always be fulfilled in the radiating near-field. $(b)$ is from \cite[eq. (24)]{LoSMIMO}. Similarly, the overall channel power is $\left\|{\bf H} \right\|_F \approx \frac{{le}^2 N_t N_r}{(4\pi d)^2}$. Then the lemma is proved.
\end{IEEEproof}

\begin{remark}
For discrete antennas, the EDoF depends on the size of MIMO, the antenna spacing, the signal frequency, and the communication distance. The aperture size of each antenna element does not affect the value of EDoF. 
\end{remark}

Note that $\lim\limits_{x \to 0} \frac{\sin^2 \left( xN_t \right)}{\sin^2 x} = {N_t}^2$, we obtain $\varepsilon _{\rm dis} = 1$ when $d \to \infty$. It means that only one data stream can be transmitted by the LoS MIMO system in the far-field.

\subsection{Capacity Analysis}

After the calculation of the EDoF, we are able to obtain the channel capacity under the fixed communication distance.

\begin{theorem}\label{theorem: d}
The capacity of the discrete MIMO system with distance $d$ has the following closed form
\begin{align}
	C_{\rm dis}(d)  = \frac{\left(N_t N_r \right)^2}{ \varphi \left( d \right)} \log_2 \left( 1+ \frac{c_0 \varphi^2\left( d \right) }{d^2 \left( N_t N_r \right)^3} \rho \right) ,
\end{align}
where $ \varphi \left( d \right) = \sum_{m_1 = 1}^{N_t} \sum_{m_2 = 1}^{N_t}  \frac{\sin^2 \left( \frac{q^2 k_0 \left(m_1 - m_2 \right) }{2d}  N_r \right)}{ \sin^2 \left( \frac{q^2 k_0 \left( m_1 - m_2  \right)}{2d}   \right)}$, $c_0 = \frac{1}{(4 \pi)^2}$, and $\rho = \frac{P}{N_0}$ is the SNR.
\end{theorem}

\begin{IEEEproof}
This theorem can be proved by combining the definition in \eqref{eq: def ecapacity} and {\bf Lemma~\ref{lemma: EDOF d}}.
\end{IEEEproof}

\subsection{Ergodic Capacity with Randomly Deployed Receivers}
In this subsection, we consider the effect of the randomly deployed receivers rather than fixing the receivers. Suppose that the centers of receivers ${\bf \Phi}_r$ are randomly deployed in a ring area with the radius $[d_1,d_2]$ in the $xy$-plane. In each time slot, the transmitter only serves a typical receiver which is randomly selected from ${\bf \Phi}_r$. In this case, the distance $d$ is a random variable, and we denote $d \sim {\cal D}$. The probability density function (PDF) of $d$ is
\begin{align}
	f_{\cal D}(d) = 2 d/{A_r},
\end{align}
where $A_r = {d_2}^2 - {d_1}^2$.

The ergodic capacity averaged over the spatial effect is defined as
\begin{align}
	{\bar C}_{\rm dis}  = \mathbb{E}_{d} \left[ C_{\rm dis} (d) \right].
\end{align}

\begin{theorem}\label{theorem: EC d}
The ergodic capacity for the discrete MIMO system is given by
\begin{align}
	{\bar C}_{\rm dis} \approx \frac{\Delta d}{A_r}\sum_{i = 1} ^{M}\omega_i g_{d}(\theta_i),
\end{align}
where $g_{d}(x) = \sqrt{1 - x^2}  \left(\frac{\Delta d}{2}x + \frac{\Delta d}{2}  \right) C_{\rm dis} \left(\frac{\Delta d}{2}x + \frac{\Delta d}{2}  \right) $, $\omega_i = \frac{\pi}{M}$, $\Delta d = d_2 - d_1$, $\theta_i = \cos \left( \frac{2i-1}{2 M} \pi \right)$, and $M$ is the parameter to ensure a complexity-accuracy trade-off.
\end{theorem}

\begin{IEEEproof}
The ergodic capacity can be calculated by ${\bar C}_{\rm dis} = \int_{d_1} ^{d_2} f_{\cal D}(x) C_{\rm dis}(x) dx$. Using the Chebyshev–Gauss quadrature, the closed-formed expression is obtained.
\end{IEEEproof}

\section{Continuous Aperture Antennas}
\subsection{Channel model}
For MIMO with continuous aperture antennas, we denote the ULA lengths of the transmitter and the receiver as $L_t$ and $L_r$, respectively. Without the loss of generality, we assume that $L_t \ge L_r$. The point ${\bf r}_t = \left(0, 0, z_t \right) \in {\cal S}_t$ on the transmitter and the point ${\bf r}_r = \left(x_r, y_r, z_r \right) \in {\cal S}_r$ on the receiver are related by the tensor Green function $G \left({\bf r}_{r}, {\bf r}_{t} \right)$, where $z_t \in [-\frac{L_t}{2}, \frac{L_t}{2}]$ and $z_r \in [-\frac{L_r}{2}, \frac{L_r}{2}]$. We suppose the vertically polarized signal and the equivalent electric currents in the $z$-direction within the transmitter. Considering that the transmitter ULA faces the receiver, the electric field at point ${\bf r}_{r}$ is $E_r \left({\bf r}\right) = \int_{{\cal S}_t} G \left({\bf r}_{r}, {\bf r} \right) J\left( {\bf r} \right) d {\bf r}$, where $J\left( {\bf r} \right) = J_z ({\bf r}) \hat{\bf u}_z$ is the transmitter current.

For ease of characterizing the correlation of the MIMO channel, we focus on the self-adjoint Hilbert-Schmidt operator $G^*G$ in the space of ${\cal S}_t$ \cite{JiaqiNF}. The kernel function is given by
\begin{align}
	K \left({\bf r}_t, {\bf r}_t' \right) = \int_{{\cal S}_r} G^* \left({\bf r}, {\bf r}_t \right) G \left({\bf r}, {\bf r}_t' \right) d {\bf r}.
\end{align}


\subsection{EDoF Analysis}
Continuous aperture MIMO can be regarded as the special case of discrete MIMO with sizeless antennas when $N_t \to \infty$ and $N_r \to \infty$ in the predefined space ${\cal S}_t$ and ${\cal S}_r$, respectively. According to \cite{GaoNF}, the EDoF is expressed as
\begin{align}
	\varepsilon _{\rm con} = \frac{ \left( \int_{{\cal S}_t} \int_{{\cal S}_r} G^* \left( {\bf r}_r, {\bf r}_t \right) G\left( {\bf r}_r, {\bf r}_t \right) d {\bf r}_r d{\bf r}_t \right)^2 }{  \int_{{\cal S}_t} \int_{{\cal S}_t} K^* \left( {\bf r}_t, {\bf r}_t' \right) K\left( {\bf r}_t, {\bf r}_t' \right) d {\bf r}_t d{\bf r}_t' }.
\end{align}

Before the calculation of EDoF, we first provide the overall channel power of MIMO, which is expressed as $\left\|{\bf H} \right\|_F = \int_{{\cal S}_t} \int_{{\cal S}_r} G^* \left( {\bf r}_r, {\bf r}_t \right) G\left( {\bf r}_r, {\bf r}_t \right) d {\bf r}_r d{\bf r}_t$.
\begin{lemma}\label{lemma: J1}
For continuous aperture MIMO, the overall channel power can be calculated as
\begin{align}\label{eq: J1}
	\left\|{\bf H} \right\|_F = c_0\frac{2L_t L_r }{d^2} - c_0v(d),
\end{align}
 where $v(d) = \ln \frac{(L_t + L_r)^2  + 4d^2}{(L_t - L_r)^2  + 4d^2}$.
\end{lemma}
\begin{IEEEproof}
	See Appendix~B.
\end{IEEEproof}

\begin{corollary}\label{corollary: cp ff}
If the distance $d \to \infty$, the overall channel power is given by
\begin{align}
	\left\|{\bf H} \right\|_F = c_0\frac{ L_t L_r }{d^2}.
\end{align}
\end{corollary}
\begin{IEEEproof}
Considering $d \to \infty$, we can calculate that $\lim \limits_{d \to \infty} v(d)  = \lim \limits_{d \to \infty} \frac{4L_t L_r}{(L_t- L_r)^2 + 4d^2} = \frac{L_t L_r}{d^2}$.
\end{IEEEproof}

\begin{lemma}\label{lemma: EDoF c}
	The EDoF of the continuous aperture MIMO system is
	\begin{align}\label{eq: EDoF c}
		\varepsilon _{\rm con} = \begin{cases} 
			u(d) & d \le \frac{2L_tL_r}{\lambda} \\ 1 & d > \frac{2L_tL_r}{\lambda},  \end{cases}
	\end{align}
	where $u(d) = \frac{\left(2L_tL_r - d^2 v(d) \right)^2}{L_tL_r \lambda d - \lambda^2 d^2/4}$ and $d_F = \frac{2L_tL_r}{\lambda}$ is the EDoF-based field boundary.
\end{lemma}

\begin{IEEEproof}
	See Appendix~C.
\end{IEEEproof}

\begin{remark}
When the distance is larger than $d_F$, the communication mode for the continuous aperture MIMO system is one. Therefore, $d_F$ can be regarded as the boundary between the near-field and the far-field.
\end{remark}
\begin{remark}
For continuous aperture antennas, the EDoF depends on the aperture size of the transceiver, the signal frequency, and the communication distance.
\end{remark}

\subsection{Capacity Analysis}
Similar to discrete MIMO, the capacity of continuous aperture MIMO is expressed by \eqref{eq: def ecapacity}. Therefore, based on the derived EDoF, we calculate the channel capacity as follows.
\begin{theorem}\label{theorem: c}
The capacity of the continuous aperture MIMO system conditioned on distance $d$ can be expressed as
\begin{align}
C_{\rm con}(d)  = \begin{cases} 
		u(d) \log_2 \left(1+ \frac{c_0 \left( \frac{L_t L_r \lambda}{d^3}  -\frac{\lambda^2}{4 d^2} \right)^2  }{ \left( \frac{2L_tL_r}{d^2} -  v(d) \right)^3 } \rho\right) & d \le d_F \\ \log_2 \left(1+ \frac{c_0  L_t L_r }{d^2 }\rho \right) & d > d_F  \end{cases}
\end{align}
\end{theorem}

\begin{IEEEproof}
This theorem is proved by substituting \eqref{eq: J1} and \eqref{eq: EDoF c} into \eqref{eq: def ecapacity}.
\end{IEEEproof}

\subsection{Ergodic Capacity with Randomly Deployed Receivers}
As in the discrete MIMO system, randomly deployed receivers are considered here. The ergodic capacity can be obtained based on {\bf Theorem \ref{theorem: c}}.
\begin{theorem}\label{theorem: EC c}
The ergodic capacity for continuous aperture MIMO is given by
\begin{align}
	{\bar C}_{\rm con} \approx \frac{\Delta d_1}{A_r}\sum_{i = 1} ^{M}\omega_i g_{c,1}(\theta_i) + \frac{\Delta d_2}{A_r}\sum_{i = 1} ^{M}\omega_i g_{c,2}(\theta_i),
\end{align}
where $\Delta d_1 = \max \left\{0,\min \{d_2,d_F\} - d_1 \right\} $, $\Delta d_2 = \max \left\{0, d_2- \max \{d_1,d_F\} \right\} $, and $g_{c,t}(x) = \sqrt{1 - x^2}  \left(\frac{\Delta d_t}{2}x + \frac{\Delta d_t}{2}  \right) C_{{\rm con},t} \left(\frac{\Delta d_t}{2}x + \frac{\Delta d_t}{2}  \right) $ for $t \in \{1,2\}$. $C_{{\rm con},1}(d)$ and $C_{{\rm con},2}(d)$ are conditional capacity when $d \le d_F$ and $d > d_F$, respectively.
\end{theorem}
\begin{IEEEproof}
The proof is similar to {\bf Theorem \ref{theorem: EC d}}.
\end{IEEEproof}

\section{Numerical Results}
In this section, we present numerical results to validate our 
analytical approach and to illustrate characteristics of the near-field. We consider $f = 28$ GHz and $N_0 = -90$ dBm. For continuous aperture antennas, the ULA lengths of the transmitter and receivers are $L_t = 100 \lambda$ and $L_r = 25 \lambda$, respectively. For a fair comparison, the overall aperture size of discrete MIMO is the same as that in the continuous case.

\begin{figure}[t!]
	\centering
	\includegraphics[width= 3.2 in]{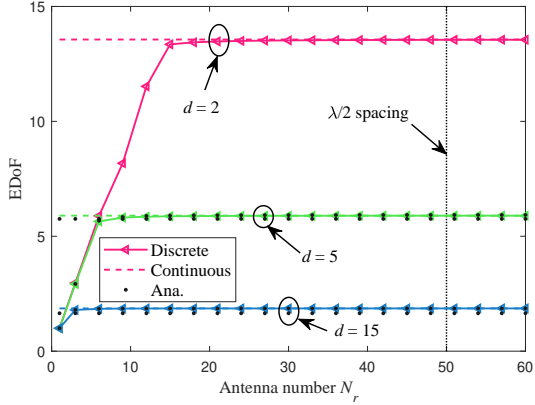}
	\caption{EDoF versus the number of antennas $N_r$. To keep the same antenna spacing $q$ at both the transmitter and the receiver, we set $N_t = 4N_r$. }\label{firure: 2}
\end{figure}

Fig. \ref{firure: 2} plots the EDoF of two categories of MIMO at different distances, which verifies {\bf Lemma \ref{lemma: EDOF d}} and {\bf Lemma \ref{lemma: EDoF c}} in the radiating near-field ($d = \{5, 15\}$ m). We find that with the increase of the antenna number, the EDoF of discrete MIMO finally converges to the continuous aperture MIMO. This convergence becomes faster when the distance $d$ is larger. We also find that in the radiating near-field, the antenna spacing for EDoF convergence is larger than the half-wavelength spacing, i.e., $q > \lambda/2$. Therefore, discrete MIMO with proper antenna spacing transmits the same number of effective data streams as continuous aperture MIMO.

\begin{figure}[t!]
	\centering
	\includegraphics[width= 3.2 in]{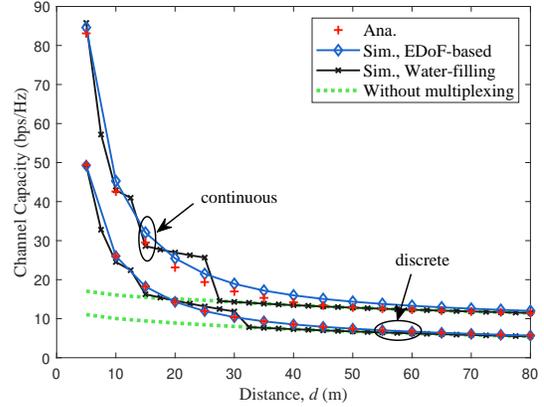}
	\caption{Channel capacity versus the communication distance $d$ with $P = 10$ dBm, $q = \lambda/2$, and $l_e = \lambda /16$. }\label{firure: 3}
\end{figure}

Fig. \ref{firure: 3} shows the channel capacity versus the distance $d$ and validates the theoretical expressions in {\bf Theorem \ref{theorem: d}} and {\bf Theorem \ref{theorem: c}}. Since the DoF (or communication mode) should be an integer in practical applications, the curves based on the optimal water-filling algorithm from \eqref{eq: capacity exact} are jagged. Our EDoF-based results can be regarded as the smoothed curves of \eqref{eq: capacity exact}. Although the EDoF-based capacity does not fit \eqref{eq: capacity exact} accurately at some distances, the expressions are tractable and capable of characterizing the average capacity over a distance. Moreover, we observe that the LoS MIMO system has a multiplexing gain of at most five times in the near-field, but the multiplexing gain attenuates to one with the increase of distance $d$. It illustrates that the spherical wave in the near-field improves the DoF of MIMO. However, in the far-field, the plane wave leads to the same channel conditions of all MIMO subchannels hence no multiplexing gain in this case.

\begin{figure}[t!]
	\centering
	\includegraphics[width= 2.9 in]{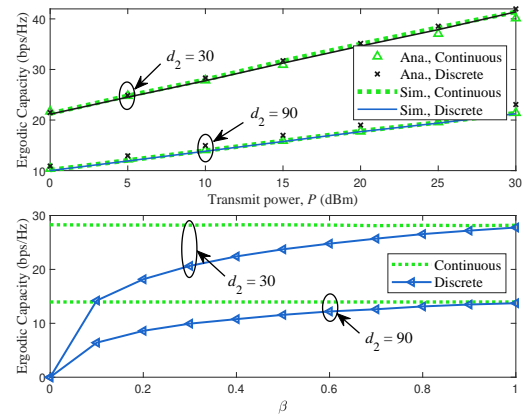}
	\caption{Ergodic capacity in different distance ranges  with $q = \lambda/2$, $d_1 = 5$m, and $M = 50$. (a) Top: ergodic capacity versus transmit power with $\beta = 1$; (b) Bottom: ergodic capacity versus $\beta$ with $P = 10$ dBm. }\label{firure: 4}
\end{figure}

Fig. \ref{firure: 4} shows the ergodic capacity in the scenario with randomly deployed receivers. We denote $\beta = l_e /q$ and consider different aperture sizes of each antenna element for the discrete MIMO receiver. Fig. \ref{firure: 4} (a) validates {\bf Theorem \ref{theorem: EC d}} and {\bf Theorem \ref{theorem: EC c}}. Due to the approximation in \eqref{eq: EDoF def} and the averaging calculation of the spatial effect, the increase of the transmit power enlarges the gap between the analytical and simulation results. We conclude that the proposed EDoF-based approach is capable of evaluating the ergodic capacity, especially in a low power regime. This approach can be utilized to analyze large-scale networks.
In Fig. \ref{firure: 4} (b), we can find that the performance of discrete MIMO converges to continuous aperture MIMO when $\beta$ increases. This can be explained that the discrete MIMO receiver becomes continuous when $\beta = 1$. Half-wavelength sampling at the MIMO transmitter, on the other hand, enables the full capability of the continuous aperture MIMO system. 
\begin{remark}
	The performance of the discrete MIMO system converges to the continuous aperture MIMO system by appropriate antenna spacing and element size, e.g., $q = \lambda/2$ and $l_e = q$. Therefore, the performance analysis on near-field MIMO can first focus on discrete MIMO, since it is easily extended to continuous aperture antennas.
\end{remark}

\section{Conclusion}
This letter proposes a tractable analytical approach to evaluate the performance of the near-field MIMO system. We have derived the closed-form expressions of the EDoF, channel capacity, and ergodic capacity, which are accurate in the low power regime. We have shown that the near-field MIMO system has a multiplexing gain even without scatter in the environment, and the multiplexing gain decreases as the communication distance increases. In addition, we have shown that the performance of continuous aperture MIMO is the upper bound of discrete MIMO. Although realizing continuous aperture MIMO is difficult in practice, conventional discrete MIMO can achieve high capacity by adjusting the antenna spacing and antenna element size.

\section*{Appendix~A: Proof of \eqref{eq: EDoF def}}
\label{Appendix:A}
\renewcommand{\theequation}{A.\arabic{equation}}
\setcounter{equation}{0}

\begin{figure}[t!]
	\centering
	\includegraphics[width= 3.2 in]{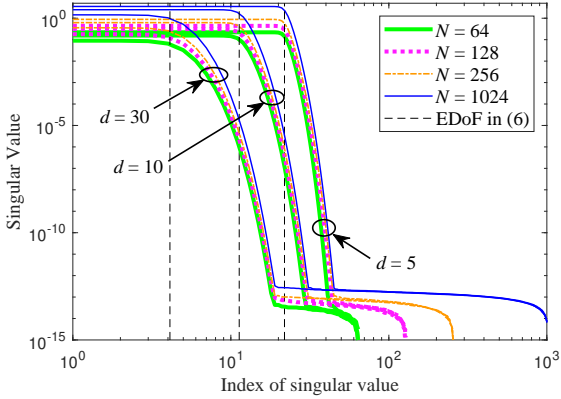}
	\caption{Singular values of LoS MIMO channel in the near-field with different numbers of antenna $N $ and communication distance $d$. }
	\label{figure: 4}
\end{figure}

For LoS MIMO, an interesting phenomenon can be found in the near-field. Considering $N_t = N_r = N$ and the length of the transmitter/receiver is $100 \lambda$. As shown in Fig. \ref{figure: 4}, those singular values of $N\times N$ MIMO obey $s_1 \approx ... \approx s_n \gg s_{n+1} > ... > s_N$, where $s_n \le s_\phi$, and the EDoF in (6) can be regarded as the boundary between $s_n$ and $s_{n+1}$. On this basis, the capacity of LoS MIMO in the low power regime can be expressed as
\begin{align}\label{eq: R1}
	C = \varepsilon \log_2 \left( 1 + \frac{\left\| {\bf H} \right\|_F \rho}{ \varepsilon^2} \right), 
\end{align}
where $\rho$ is the transmit SNR. From \eqref{eq: R1} we observe that the channel capacity of MIMO can be expressed as the sum capacity of $\varepsilon$ identical SISO channels in the low power regime.

On the other hand, exact capacity can be expressed as \eqref{eq: R1} with an accurate $\varepsilon$. Now let us prove that the expression (6) converges to the exact value when the bit SNR $\frac{E_b}{N_0} \to {\frac{E_b}{N_0}}_{\min}$, where ${\frac{E_b}{N_0}}_{\min}$ is the minimum bit SNR required for reliable communication.
Using the fact $\rho = \frac{E_b}{N_0} C$ and expression \eqref{eq: R1}, the relationship between capacity $C$ and bit SNR $\frac{E_b}{N_0}$ can be given by
\begin{align}\label{eq: R2}
	\frac{E_b}{N_0} = \frac{2^{C/\varepsilon} - 1}{ \left( C/\varepsilon \right) \left( \left\| {\bf H} \right\|_F /\varepsilon \right) }.
\end{align}

According to references \cite{defEDoF,verdu}, in the low power regime, the capacity can be expressed as
\begin{align}\label{eq: R3}
	C = {\cal S}\left(\log_2 \frac{E_b}{N_0} - \log_2 { \frac{E_b}{N_0}}_{\min} \right),
\end{align}
where ${\cal S} = \frac{ \left( {\rm tr} \left( {\bf R} \right) \right)^2}{ {\rm tr} \left(  {\bf R}^2 \right)}$ and ${ \frac{E_b}{N_0}}_{\min} = \frac{\varepsilon\ln 2}{{\rm tr} \left( {\bf R} \right)} = \frac{\varepsilon\ln 2}{\left\| {\bf H} \right\|_F}$. We denote $\Delta \Omega = \frac{E_b}{N_0} / {\frac{E_b}{N_0}}_{\min}$. By substituting \eqref{eq: R3} into \eqref{eq: R2}, we obtain
\begin{align}
	\frac{E_b}{N_0} = \frac{{\Delta \Omega}^\frac{{\cal S}}{\varepsilon} - 1}{  \left\| {\bf H} \right\|_F {\cal S} \log_2 \Delta\Omega  }\varepsilon^2. 
\end{align}

When $\frac{E_b}{N_0} \to {\frac{E_b}{N_0}}_{\min}$, the below equation holds
\begin{align}
	\lim\limits_{\Delta \Omega \to 1} \Delta \Omega  = \lim\limits_{\Delta \Omega \to 1} \frac{ {\Delta \Omega}^\frac{{\cal S}}{\varepsilon} - 1}{ \frac{{\cal S}}{\varepsilon} \ln 2  \log_2  \Delta\Omega  } = 1. 
\end{align}

Therefore, we have $\varepsilon = {\cal S} = \frac{ \left( {\rm tr} \left( {\bf R} \right) \right)^2}{ {\rm tr} \left(  {\bf R}^2 \right)}$ when $\frac{E_b}{N_0} \to {\frac{E_b}{N_0}}_{\min}$. The convergence of (6) is proved.

Based on the discussion above, the EDoF $\varepsilon$ can be approximated by (6) in the low power regime.

\section*{Appendix~B: Proof of Lemma~\ref{lemma: J1}}
\label{Appendix:B}
\renewcommand{\theequation}{B.\arabic{equation}}
\setcounter{equation}{0}

Based on the definition of the Green's function, $\left\|{\bf H} \right\|_F$ can be calculated as
\begin{align}\label{eq: A.1}
	\left\|{\bf H} \right\|_F &= \frac{1}{(4\pi)^2 } \int_{-\frac{L_t}{2}}^{\frac{L_t}{2}} \int_{-\frac{L_r}{2}}^{\frac{L_r}{2}} \frac{1}{d^2 + \left( z_t - z_r \right)^2} d z_r dz_t \nonumber \\
	& = \frac{L_t L_r}{(4\pi)^2 } \int_{0}^{\frac{L_t+ L_r}{2}} \frac{f_{\Delta z}(z)}{d^2 + z^2} dz,
\end{align}
where $f_{\Delta z}(z)$ is the PDF of the distance $\Delta z = |z_t -z_r|$. Since ${\bf r}_t$ and ${\bf r}_r$ are uniformly distributed on ${\cal S}_t$ and ${\cal S}_r$, respectively, the $f_{\Delta z}(z)$ is given by
\begin{align}\label{eq: A.2}
	f_{\Delta z}(z) = \begin{cases} 
		\frac{2}{L_t} & z \in [0, u_1] \\ \frac{L_t + L_r -2z}{L_t L_r} & z \in [u_1, u_2],  \end{cases} 
\end{align}
where $u_1 = \frac{L_t - L_r}{2}$ and $u_2 = \frac{L_t + L_r}{2}$. By substituting \eqref{eq: A.2} into \eqref{eq: A.1}, we have 
\begin{align}
	\left\|{\bf H} \right\|_F = &\frac{1}{(4\pi)^2 } \underbrace{\left(  \int_{0}^{u_1} \frac{L_t L_r}{d^2 + z^2} \frac{2}{L_t} dz + \int_{u_1}^{u_2} \frac{L_t + L_r}{d^2 + z^2} dz \right)}_{f_1} \nonumber \\
	&- \frac{1}{(4\pi)^2 } \int_{u_1}^{u_2} \frac{2z}{d^2 + z^2}  dz ,  
\end{align}
For the part $f_1$, it can be calculated as follows
\begin{align}
	f_1 &\overset{(a)}{=} \frac{2L_r }{d} \tan^{-1} \frac{u_1}{d} + \frac {L_t +L_r}{d}  \left( \tan^{-1} \frac{u_2}{d} -\tan^{-1} \frac{u_1}{d} \right)  \nonumber \\
	& \overset{(b)}{\approx} \frac{(L_r-L_t)u_1 }{d^2} + \frac{(L_r+L_t)u_2}{d^2},
\end{align}
where $(a)$ is from \cite[eq. (2.124.1)]{Intetable}. $(b)$ is obtained by using the Taylor expansion $\tan^{-1} x = x + o(x^3)$.

\section*{Appendix~C: Proof of Lemma~\ref{lemma: EDoF c}}
\label{Appendix:C}
\renewcommand{\theequation}{C.\arabic{equation}}
\setcounter{equation}{0}

We denote $J_0 = \int_{{\cal S}_t} \int_{{\cal S}_t} K^* \left( {\bf r}_t, {\bf r}_t' \right) K\left( {\bf r}_t, {\bf r}_t' \right) d {\bf r}_t d{\bf r}_t'$ for simplicity, which is the overall kernel power. To calculate $J_0$, we first focus on the kernel function $K\left( {\bf r}_t, {\bf r}_t' \right)$ expressed as
\begin{align}\label{eq: kernal}
	K\left( {\bf r}_t, {\bf r}_t' \right) = \int_{- \frac{L_r}{2}} ^{ \frac{L_r}{2}}  \frac{ c_0 F_T^*(z_t) F_T(z_t')}{\sqrt{ ( d^2 + |z_t - z|^2) (d^2 + |z_t' - z|^2)}} \nonumber \\
	\times \exp \left( - jk_0\frac{z \left( z_t - z_t' \right) }{d} \right)  d z,
\end{align}
where $F_T(z) = \exp \left( -jk_0\frac{z^2}{2d} \right)$ is the focusing phase-factor function. When ${\bf r}_t$ and ${\bf r}_t'$ are far away from each other, the oscillatory of the phase term in \eqref{eq: kernal} leads the value of kernel function to zero. Specifically, when
\begin{align}\label{eq: volume ele}
	\left| \frac { k_0L_r \left(z_t - z_t' \right)}{d}\right| \ge \pi,
\end{align}
we have $K\left( {\bf r}_t, {\bf r}_t' \right) = 0$ \cite{Miller}. Otherwise, ${\bf r}_t$ and ${\bf r}_t'$ have strong correlation. The kernel power can be approximated as
\begin{align}
	\left|K\left( {\bf r}_t, {\bf r}_t' \right) \right|^2 \approx \left(  \frac{ c_0 L_r }{ d^2 + f(z_t, z_t')^2} \right)^2,
\end{align}
where $f(z_t, z_t') = \frac{z_t + z_t'}{2}$. Then we are able to express $J_0$ as a surface integral
\begin{align}
	J_0 = \iint_{{\cal S}_{\Delta_t}} \left(  \frac{ c_0 L_r }{ d^2 + g(x, y)^2} \right)^2 dx dy,
\end{align}
where surface ${\cal S}_{\Delta_t}: \left\{ \left| y-x \right| \le \frac{\Delta_t}{2} \right\} \cap \left\{ \left| x \right| \le \frac{L_t}{2} \right\} \cap \left\{ \left|y \right| \le \frac{L_t}{2} \right\} $. 
From \eqref{eq: volume ele}, we have $\Delta_t = 2 \max \left|z_t - z_t' \right| = \frac{\lambda d}{L_r}$. When $d \le \frac{2L_tL_r}{\lambda}$, the expression of $J_0$ can be further simplified as
\begin{align}
	J_0 = 2 \Delta_t {c_0}^2 {L_r}^2  \int_{0} ^{u_0}   \frac{ 1}{\left( d^2 + z^2 \right)^2}  dz \nonumber \\
	+ 2 {c_0}^2 {L_r}^2  \int_{u_0} ^{\frac{L_t}{2}}   \frac{ 2L_t - 4z}{\left( d^2 + z^2 \right)^2}  dz,
\end{align}
where $u_0 = \frac{L_t}{2} - \frac{\Delta_t}{4}$. By employing \cite[eq. (2.148.4)]{Intetable} and the Taylor expansion of $\tan^{-1} x$, we obtain the closed-form expression after some tedious derivations
\begin{align}
	J_0 = {c_0}^2 \frac{L_tL_r \lambda}{d^3} - {c_0}^2 \frac{\lambda^2}{4d^2}.
\end{align}

Similarly, when $d > \frac{2L_tL_r}{\lambda}$, we have
\begin{align}
	J_0 &= 2 {c_0}^2 {L_r}^2  \int_{0} ^{\frac{L_t}{2}}   \frac{ 2L_t - 4z}{\left( d^2 + z^2 \right)^2}  dz  = {c_0}^2 \frac{ \left( L_t L_r \right)^2}{d^4}.
\end{align}

Afterwards, using the results in {\bf Lemma \ref{lemma: J1}} and {\bf Corollary \ref{corollary: cp ff}}, the lemma is proved.

\vspace{-0.1 cm}
\bibliographystyle{IEEEtran}
\bibliography{reference}


\end{document}